\documentclass{ws-procs9x6}
\usepackage{color,rotate}
 
\begin{document}
\newcommand{\be}{\begin{eqnarray}}
\newcommand{\ee}{\end{eqnarray}}
\newcommand\del{\partial}
\newcommand{\mat} {\left ( \begin{array}{cc}}
\newcommand{\emat} { \end{array}\right )}
\newcommand{\matt}{\left ( \begin{array}{ccc}}
\newcommand{\ematt}{\end{array} \right )}
\newcommand{\matf}{\left ( \begin{array}{cccc}}
\newcommand{\ematf}{\end{array} \right )}
\newcommand{\vect}{\left ( \begin{array}{c}}
\newcommand{\evect}{\end{array} \right )}

\title{QCD Dirac Spectra and the Toda Lattice\footnote{\uppercase{T}his 
work is supported in part by \uppercase{US DOE} grant
\uppercase{N}o. \uppercase{DE}-\uppercase{FG}-88\uppercase{ER}40388.}} 

\author{K. SPLITTORFF}

\address{Nordita, 
Blegdamsvej 17, \\
DK-2100, Copenhagen, Denmark\\
split@alf.nbi.dk}

\author{J.~J.~M. VERBAARSCHOT}

\address{Department of Physics and Astronomy,\\
SUNY at Stony Brook, Stony Brook, NY\,11794, US\\
jacobus.verbaarschot@stonybrook.edu}

\maketitle

\abstracts{We discuss the spectrum of the QCD Dirac operator both
at zero and at nonzero baryon chemical potential. We show that,
in the ergodic domain of QCD, the Dirac spectrum can be obtained
from the replica limit of a Toda lattice equation. At zero chemical
potential this method explains the factorization of known
results into compact and noncompact integrals, and at nonzero chemical
potential it allows us to derive the previously unknown
microscopic spectral density.}

\section{Introduction}
Because of the spontaneous breaking of chiral symmetry and confinement, 
QCD at low energy is a theory of weakly interacting Goldstone bosons.
In the spontaneously broken phase,  
the  QCD partition function is a nonanalytic function of the
quark mass with a  chiral condensate that is discontinuous as the
quark mass crosses the eigenvalue axis of the QCD Dirac operator. 
The strength of this discontinuity is proportional to the eigenvalue density,
a relation known as the Banks-Casher formula \cite{BC}.

A theory  with spontaneously broken chiral symmetry that is much simpler than
QCD is chiral Random Matrix Theory  \cite{SV}. This is a theory with the
global flavor symmetries of QCD in which the
matrix elements of the Dirac operator are replaced by random numbers.
Although, this theory is  zero dimensional,
chiral symmetry is broken spontaneously in the limit of
infinitely large matrices, 
and the mass of the non-Goldstone 
modes diverges in the limit $N \to \infty$,
where $N$ is the matrix size. Therefore, in the
thermodynamic 
limit, $N\to\infty$, chiral Random Matrix theory reduces to a theory 
of Goldstone bosons for which, in the limit of small quark masses, the
Lagrangian is just the mass term of the chiral Lagrangian. 
This is the main reason why
Random Matrix Theories have been so successful in this context.

One of the questions we have been asking is whether we can identify
a parameter domain where QCD and chiral Random Matrix Theory reduce 
to the same theory of Goldstone bosons.
The affirmative answer to this question is that this is the case if 
the Compton wavelength of the Goldstone bosons is much larger than 
the linear size $L$ of the box.
This requires that the quark masses
$m_f \ll F^2/(\Sigma L^2)$, which is an unphysical domain of QCD, so that 
the kinetic term of the chiral Lagrangian can be ignored \cite{GL,LS}.
($\Sigma$ is the chiral condesate and $F$ is the pion decay constant.) 
However, even for realistic quark masses we can identify a parameter
domain where QCD and chiral Random Matrix Theory behave the same, namely 
the domain
where eigenvalues of the Dirac operator $\lambda \ll F^2/(\Sigma L^2)$
which is known as the ergodic domain \cite{Vplb}.
The reason
is that the generating function of the Dirac spectrum is a QCD-like
partition function with additional ghost quarks with mass $\lambda$.
The condition for the validity of the Random Matrix Theory description
of the Dirac spectrum is
then that the Compton wavelength of Goldstone bosons composed out of
ghost quarks
is much larger than the size of the box. Indeed, such behavior has
been observed in numerous lattice QCD simulations (as discussed in 
detail elsewhere \cite{review,paullattice}).

At nonzero chemical potential the eigenvalues of the Dirac operator
are scattered in the complex plane. It has been shown that the
Dirac spectrum remains in the ergodic domain if the inverse
chemical potential is much larger than the size of the box \cite{TV}. 
In this domain the Dirac spectrum can be described by a chiral
Random Matrix Model that has been extended with a chemical
potential \cite{misha,HOV}. However, until recently, 
this random matrix model had only been
 solved at the mean field level \cite{misha}. The standard
methods to derive nonperturbative results such as the supersymmetric
method \cite{Efetov}
and  complex orthogonal polynomial methods \cite{french,Fyodorov,akemann} 
have not been successful in this case. So far the supersymmetric method failed
because of technical problems in calculating the graded integrals, 
while the method of
complex orthogonal polynomials failed because of the absence
of an eigenvalue
representation at non-zero chemical potential.  
The mean field analysis of the random matrix model 
was performed using the replica trick
\cite{EA} 
which was widely believed to only work  for the derivation of
perturbative results \cite{critique}. However, it was shown 
recently that if a family of partition functions has certain integrability
properties it is possible to obtain exact nonperturbative results
by means of the replica trick \cite{kanzieper02,SplitVerb1}. We
will show that the ergodic limit of the 
phase quenched QCD partition function
at  nonzero chemical potential has the required integrability properties
\cite{SplitVerb2}
so that the hierarchy of phase quenched partition functions with
a different number of flavors are related by a recursion relation
\cite{SplitVerb2}
which is known as the Toda lattice equation. The spectral density
is then obtained from the replica limit of this 
Toda lattice equation.

In the first part of this lecture we discuss the Dirac spectrum
at zero chemical potential. We show that in the ergodic domain
the spectral density can be obtained from the replica limit
of the Toda lattice equation. This result explains a factorization
property of the resolvent. In the second half of this lecture
we study the quenched Dirac spectrum at nonzero chemical potential.
Using the replica limit of the Toda lattice equation we derive the
analytical result for the microscopic spectral density
in the ergodic domain.

\section{The Dirac Spectrum in QCD}

The eigenvalues $\{\lambda_k\}$ of the anti-hermitian 
Dirac operator are determined by the eigenvalue equation
\be
D\phi_k = i\lambda_k \phi_k.
\ee
Because of the axial symmetry the nonzero eigenvalues occur
in pairs $\pm \lambda_k$. The number of zero eigenvalues is
almost always equal to the topological charge of the gauge field
configuration. The average spectral density, defined by
\be
\rho(\lambda) = \left\langle \sum_k \delta(\lambda-\lambda_k) \right\rangle, 
\ee
can be obtained from the discontinuity of the average 
resolvent
\be
G(z) = \left\langle \sum_k \frac 1{z+i\lambda_k} \right\rangle,
\label{G} 
\ee
where, in both cases, the average is over gauge field configurations
distributed according to the QCD action.
\begin{figure}[ht] 
\centerline{\hspace*{2cm}\epsfysize=6cm
\epsfbox{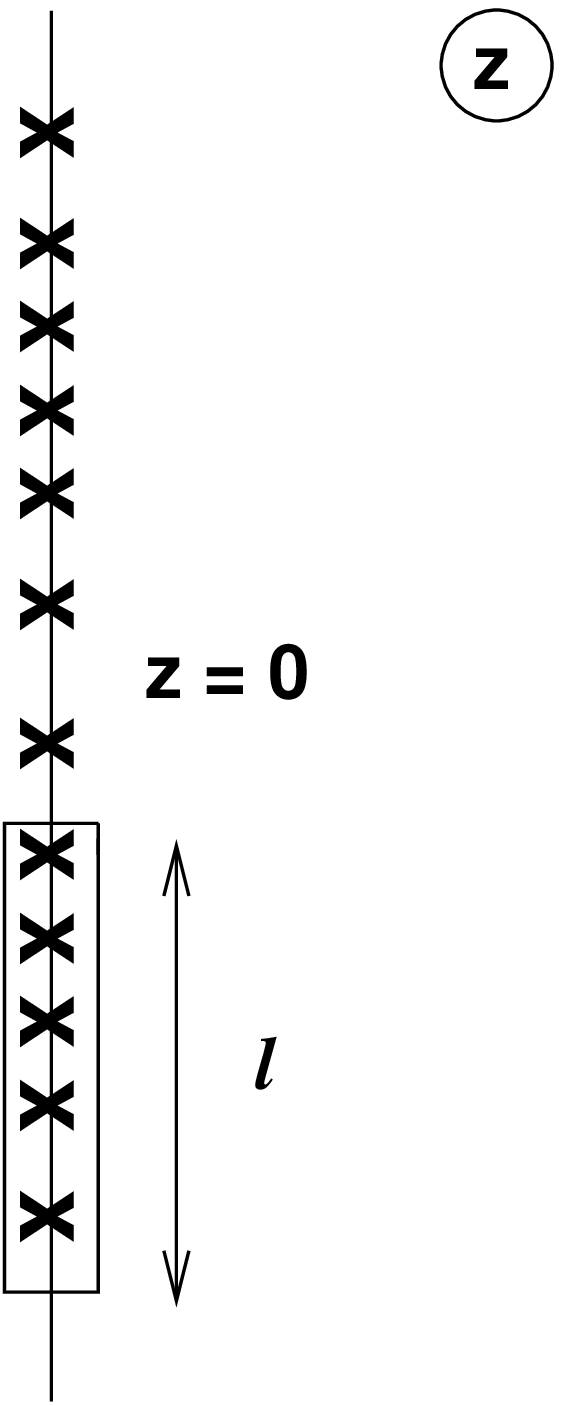}\hspace*{0.5cm}
\epsfxsize=8cm\epsfbox{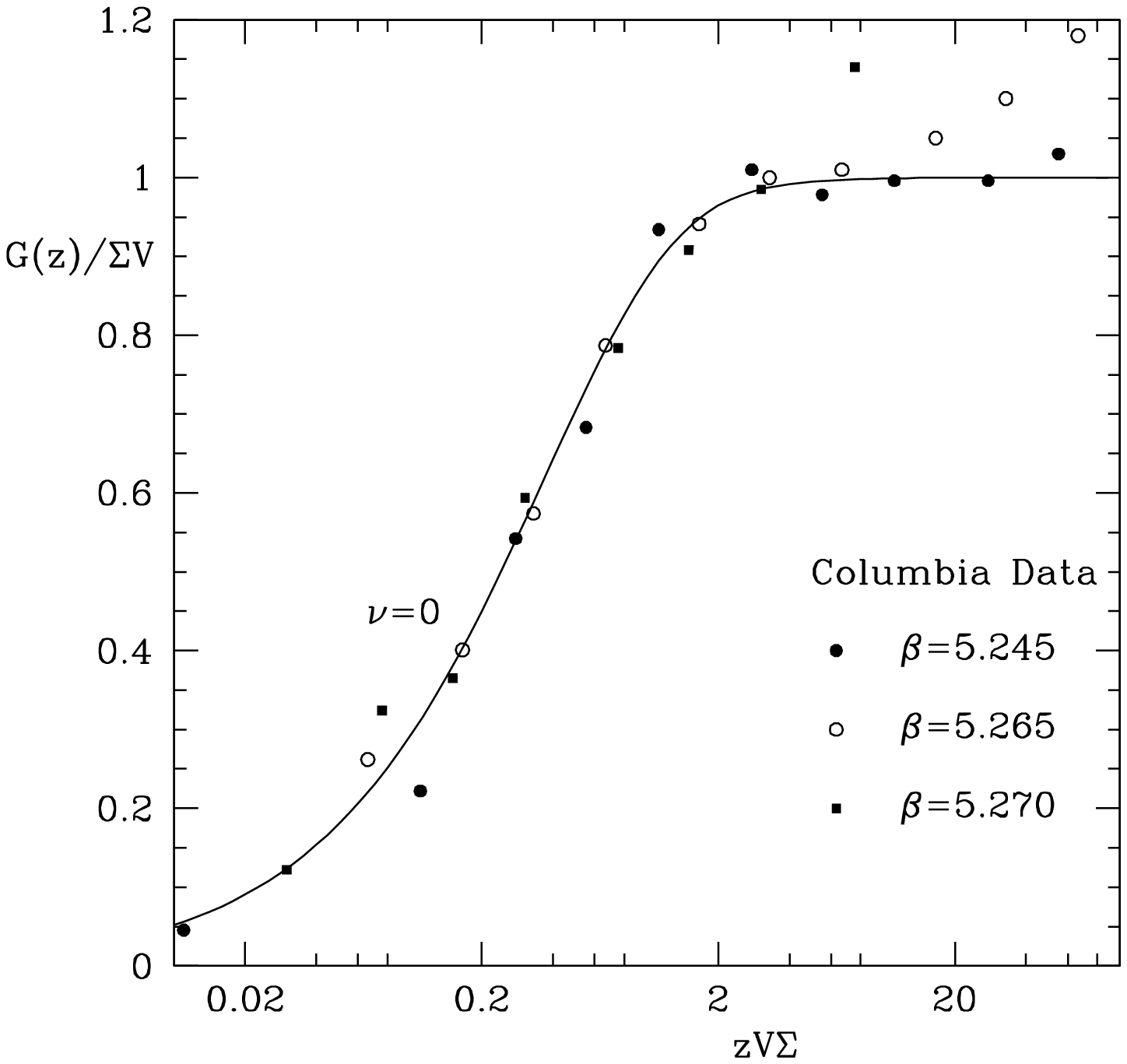} }   
\caption{ A typical Dirac spectrum (left) and the average resolvent $G(z)$
in units of $\Sigma V$ compared with lattice QCD data.
\label{fig1}}
\end{figure}
This can be seen by considering the rectangular contour in Fig. 1. Assuming
that the average spectral density does not vary substantially along this
contour, the average total number of eigenvalues inside this 
contour is $\rho(\lambda) l$, where $\lambda$ is a point on the imaginary
axis inside this contour. Therefore, if
we integrate the resolvent along this contour, we obtain
\be 
\oint G(z) = il(G(i\lambda+\epsilon) - G(i\lambda -\epsilon))
=2 \pi i \rho(\lambda) l.
\ee
Using the symmetry of the spectrum we find \cite{BC}
\be
{\rm Re}[G(i\lambda+\epsilon)] = \pi \rho(\lambda).
\ee
In QCD, the chiral condensate defined by $\Sigma = G(\epsilon)/V$
where $V$ is the volume of space time, is nonzero because of the
spontaneous breaking of chiral symmetry. The average level spacing
near zero is therefore given by
\be 
\Delta  = \frac 1{\rho(0)} = \frac{\pi}{\Sigma V},
\ee
which is also the scale of the smallest nonzero eigenvalue.
Asymptotically, for large $\lambda$, the Dirac eigenvalues
approach the spectrum of a free Dirac operator with eigenvalue
density given by $\sim V\lambda^3$.

Because the eigenvalues near zero are spaced as $1/\Sigma V$ it makes
sense to introduce the microscopic spectral density \cite{SV}
\be
\rho_s(\lambda) = \lim_{V\to \infty } \frac 1{V\Sigma}\,
\rho\left ( \frac \lambda{V\Sigma}\right ).
\ee
In the ergodic domain of QCD this is a universal function that can also
be derived from chiral Random Matrix Theory.

\subsection{Ergodic Domain of QCD}

The low-energy limit of QCD for $N_f$ fermionic flavors
is described by a theory of weakly interacting
Goldstone bosons parametrized by the unitary matrix $U(N_f)$. If their mass 
\be
m_\pi^2 = \frac{2m \Sigma}{F^2} \ll \frac {\pi^2}{L^2},
\ee
(with $F$ the pion decay constant)
the kinetic term of the chiral Lagrangian factorizes from the partition
function and the mass dependence of the partition function in the sector of
topological charge $\nu$ is given by \cite{GLeps}
\be
Z^\nu_{N_f}(M) = \int_{U(N_f)}{ \det}^\nu U e^{ \frac 12\Sigma V {\rm Tr}
[M U^{-1} + M^\dagger U]},
\label{zfer}
\ee
where the quark mass matrix is defined by $M={\rm diag}(m_1, \cdots, m_{N_f})$.
This is the ergodic domain of QCD.

As we will
explain below partition functions with bosonic quarks are essential for
obtaining the resolvent and average spectral density of the Dirac operator. 
For bosonic quarks, the Goldstone bosons cannot be parameterized by 
a unitary matrix. The reason is that symmetry transformations have
to be consistent with the convergence of the bosonic integrals.
Let us consider the case of one bosonic flavor. Then
\be
{\det}^{-1} \mat m & id\\ id^\dagger & m \emat =\frac 1{\pi^2}
\int d^2\phi_1 d^2\phi_2 \exp \left [-
\vect \phi_1^* \\ \phi_2^* \evect^T \mat m & id\\ id^\dagger & m \emat
\vect \phi_1 \\ \phi_2 \evect \right ],
\label{detinv}
\ee
so that the exponent is purely imaginary for $m=0$ and convergent for
${\rm Re}(m) > 0$.

The most general flavor transformation of the action in (\ref{detinv}) 
is a $Gl(2)$ transformation that  can be parameterized as 
\be
U = e^H V \quad {\rm with } \quad H^\dagger = H \quad {\rm and } \quad
V V^\dagger =1 .
\ee
For $U$ to be a symmetry transformation for $m = 0$  we require that
\be
U^\dagger \mat 0 & id \\ id^\dagger & 0 \emat U =
 \mat 0 & id \\ id^\dagger & 0 \emat,
\ee
so that $H$ has to be a multiple of $\sigma_3$ and $V$ has to be a 
multiple of the identity. 
The $V$ part of $U$  is not broken by 
the mass term and is thus a vector symmetry. Only the symmetry
transformation $\exp(s\sigma_3)$ is broken by the mass term
 so that the axial transformations can be parameterized by
\be
U = \mat e^s & 0 \\ 0 & e^{-s} \emat \qquad {\rm with} 
\quad s \in \langle -\infty, \infty \rangle.
\ee
For $N_f$ bosonic flavors the axial transformations are parameterized by
\be
U = \mat e^H & 0 \\ 0 & e^{-H} \emat \qquad {\rm with} 
\quad H^\dagger  = H.
\ee
The Goldstone manifold is thus $Gl(N_f)/U(N_f)$.
In the domain were the kinetic term of the chiral Lagrangian can
be ignored, the effective bosonic 
partition function (indicated by the subscript $-N_f$) in the sector
of topological charge $\nu$ is given by \cite{OTV,DOTV,DV1}
\be
Z^\nu_{-N_f}(m_f) = \int_{Gl(N_f)/U(N_f)}{ \det}^\nu U e^{ \frac 12\Sigma V 
{\rm Tr}[M U^{-1} + M^\dagger U]}.
\label{zbos}
\ee

The resolvent can be obtained from a supersymmetric generating function
that contains one additional fermionic ghost quark and one additional
bosonic ghost quark,
\be
G(z) = \partial_z \left . \left\langle \frac{\det(D+z)}{\det(D+z')}
\right\rangle 
\right |_{z'=z}.
\ee
Therefore the low energy limit of this generating function contains
additional ghost Goldstone bosons and fermions
with mass given
by $2z \Sigma/F^2$. For 
$z\ll F^2/\Sigma L^2$ the $z$-dependence of this generating function
is given by  \cite{OTV,DOTV}
\be
Z_{N_f+1;-1}^\nu(M)=\int_{\hat{Gl}(N_f+1|1)}
{ \det}^\nu U e^{ \frac 12\Sigma V {\rm STr}
[M U^{-1} + M^\dagger U]},
\ee
with $M = {\rm diag}(m_1,\cdots, m_{N_f}, z, z')$  and 
$\hat{Gl}(N_f+1|1)$ are super-matrices with a unitary $U(N_f+1)$
upper left block, a $Gl(1)/U(1)$ lower right block and 
Grassmann valued matrix elements 
elsewhere. The number of QCD Dirac eigenvalues that is described
by this partition function is of the order 
\be
\frac{F^2}{\Sigma L^2\Delta } = F^2 L^2.
\ee
This number increases linearly in $N_c$ for $N_c\to \infty$ which
was recently confirmed by lattice simulations \cite{neuberger}.

An alternative to the supersymmetric method is to use the replica trick 
to calculate the resolvent. It comes in two different versions:
the fermionic replica trick defined by
\be
G(z) = \lim_{N_f \to 0} \frac 1{N_f} \log Z_{N_f}^\nu(z),
\label{grepfer}
\ee
and the bosonic replica trick defined by
\be
G(z) = \lim_{N_f \to 0} \frac 1{-N_f} \log Z_{-N_f}^\nu(z).
\label{grepbos}
\ee
If we take the replica limit of the fermionic (\ref{grepfer}) or bosonic
(\ref{grepbos}) partition functions directly, 
we will obtain a result that differs from the
supersymmetric calculation \cite{critique}. For almost two decades there
were no methods to do reliable nonperturbative calculations with the
replica trick. In the next section we will
show that these problems with the replica trick can be avoided if the take
the replica limit of the Toda lattice equation.

\subsection{Toda Lattice Equation}

We now consider bosonic and fermionic partition functions with all
masses equal to $z$ which only depend on the combination, cf. (\ref{zfer})
and (\ref{zbos}), 
\be
x = z \Sigma V.
\ee
The unitary integral in 
the fermionic partition function (\ref{zfer})
can be evaluated by decomposing
$U = V {\rm diag}(e^{i\theta_k} ) V^\dagger $ and choosing the $\{\theta_k\}$
and $V$ as new integration variables.
By expanding the Jacobian of this transformation, given by
$\prod_{k<l} |\exp(i\theta_k) - \exp(i \theta_l)|^2$, the different terms
factorize into products of modified Bessel functions which can be
combined again into a single determinant. The final result is given by
\cite{tan,KSS}
\be
Z_{N_f}^\nu(x) =  \det\left[ I_{\nu+k-l}(x)\right]_{k,l = 1,\cdots, N_f}.
\ee
By using recursion relations for the Bessel functions, this 
result can be rewritten as 
\be
Z_{N_f}^\nu(x) = \frac 1{x^{N_f(N_f-1)}}
 \det \left[(x\del_x)^{k+l}I_{\nu}(x)\right]_{k,l = 0,\cdots, N_f-1}.
\label{zfereig} 
\ee

Next we use
the Sylvester identity \cite{Sylvester,forrester}
which is valid for determinant of an arbitrary
matrix $A$. It is given by
\be
C_{ij} C_{pq} - C_{iq}C_{pj} = \det (A)C_{ij;pq},
\ee
where the $C_{ij}$ are cofactors of the matrix $A$ and
the $C_{ij;pq}$ are double cofactors. By applying this identity
to the determinant in (\ref{zfereig})
for $i=j=N_f-1$ and $p=q=N_f$,  we easily derive the Toda lattice 
equation \cite{dijkgraaf,Kharchev,mmm}
\be
(x\del_x)^2 \log Z_{N_f}^\nu(x) =  2N_f x^2 
\frac{Z_{N_f+1}^\nu(x) Z_{N_f-1}^\nu(x)}{[Z^\nu_{N_f}(x)]^2}.
\label{toda}
\ee
This equation has also been derived as a consistency condition
for QCD partition functions \cite{paulconsist}. It also occurs
in other application such as for example in self-dual Chern-Simons
theory \cite{dunne}. 

In the case of bosonic quarks (\ref{zbos}) the positive definite matrix is
diagonalized as $U = V {\rm diag}(e^{s_k}) V^\dagger$. Choosing
the $\{s_k\}$ and $V$ as new integration variables, we obtain
\cite{DV1}
after expanding the   Jacobian 
given by $\prod_{k<l} (\exp(s_k) -\exp(s_l))(\exp(-s_k) - \exp(-s_l))$,
\be
Z^\nu_{-N_f}(x) =  \det \left[K_{\nu+k-l}(x)\right]_{k,l = 1,\cdots, N_f}.
\ee
As in the fermionic case, this result can be written in the
form of a $\tau$ function
\be
Z^\nu_{-N_f}(x) = \frac 1{x^{N_f(N_f-1)}}
\det \left[(x\del_x)^{k+l}K_{\nu}(x)\right]_{k,l = 0,\cdots, N_f-1},
\label{zboseig}
\ee
where we have used that $I_\nu$ and $(-1)^\nu K_\nu$ satisfy the
same recursion relations. Therefore, the bosonic partition function
also satisfies the Toda lattice equation (\ref{toda}).
 
The two semi-infinite hierarchies are connected by
\be
\lim_{N_f  \to 0} \frac{1}{N_f}(x\del_x)^2 \log Z_{N_f}^\nu(x).
\ee
By extending the Toda lattice hierarchy to include an additional
spectator boson, it can be shown that \cite{SplitVerb3}
\be
\lim_{N_f  \to 0} \frac{1}{N_f}(x\del_x)^2 \log Z_{N_f}^\nu(x)&=& 
\lim_{y\to x} x\del_x(x\del_x + y \del_y) \log Z^\nu_{1,-1}(x|y)\nonumber\\
&=&x\del_x \lim_{y\to x} x\del_x \log Z^\nu_{1,-1}(x|y)
\nonumber \\ &=& x\del_x x G(x).
\ee
Taking the replica limit of the Toda lattice equation (\ref{toda}) we 
thus obtain
the identity 
\be
x\del_x x G(x) = 2 x^2 Z_1^\nu(x) Z_{-1}^\nu (x),
\ee
which explains this factorization property. 
In the same way we can show the factorization of the
susceptibility into a bosonic and a fermionic partition function
\cite{SplitVerb2}.

Inserting the expressions for $Z_1$ and $Z_{-1}$ we find
\be
G(x) = x(K_\nu(x)I_\nu(x) + K_{\nu-1}(x) I_{\nu+1}(x)) +\frac{\nu}{x}.
\ee
For $\nu = 0$ this result is shown as $G(x)/\Sigma V$
by the solid curve
in the right figure of Fig. \ref{fig1}.
Agreement with lattice data \cite{CC} is found in the
ergodic domain of QCD.

\section{Dirac Spectrum at Nonzero Chemical Potential}

At nonzero baryon chemical potential the Dirac operator is
modified according to
\be
D \rightarrow D + \mu \gamma_0 .
\ee
This Dirac operator does not have any hermiticity properties
and its eigenvalues are scattered in the complex 
plane \cite{all,Markum,maria,hands,tilo}.
For small $\mu$ we expect that the width of the cloud of
eigenvalues \cite{all} $\sim \mu^2$. The average spectral density is given by
\be 
\rho(\lambda) = \left\langle \sum_k \delta^2(\lambda- \lambda_k) \right\rangle,
\ee
and the average resolvent is defined as usual by (\ref{G}).
They are related by
\be
\del_{z^*} G(z)|_{z=\lambda} = \pi \rho(\lambda).
\ee
The quenched spectral density is therefore given by the replica limit
\cite{Girko,Goksch,misha} 
\be
\rho(z,z^*) = \lim_{n \to 0} \frac 1{\pi n}\del_z \del_{z^*} \log Z_n(z,z^*),
\label{rhoreplica}
\ee
with generating function given by (note that $n$ counts pairs of quarks)
\be
Z_n(z,z^*) = \left\langle {\det}^n(D + \mu\gamma_0 +z)
{\det}^n(-D + \mu\gamma_0 +z^*)\right\rangle. 
\ee

The low-energy limit of this generating function is a chiral
Lagrangian which is determined by its global symmetries and 
transformation properties. 
By writing the product of the two determinants as \cite{TV}
\be
{\det}(D + \mu\gamma_0 +z)
{\det}(-D + \mu\gamma_0 +z^*)
= 
\left| \begin{array}{cccc} id+\mu &0 &z& 0 \\
                            0& id-\mu & 0& z^*\\
                 z & 0 & id^\dagger+\mu  & 0 \\
                  0 & z^* & 0& id^\dagger -\mu \\ \end{array} \right |
\nonumber \\
\ee
we observe that the $U(2) \times U(2)$ flavor symmetry is broken
by the chemical potential term and the mass term. Invariance is recovered
by transforming the mass term as in the case of zero chemical potential
and the chemical potential term by a local gauge transformation. In
the domain of $\mu$ and $z$ where we can neglect the kinetic terms, 
the partition function is given by \cite{TV,splitverb2}
\be
Z_n(z,z^*) = \int_{U(2n)} dU e^{-\frac{F^2\mu^2 V}4 {\rm Tr}[U,B][U^{-1},B]
+\frac {\Sigma V}2{\rm Tr}M(U+U^{-1})},
\label{genmu}
\ee
where
\be\label{BM}
B= \mat {\bf 1}_n &0 \\ 0 & - {\bf 1}_n \emat, \qquad
M = \mat z{\bf 1}_n &0 \\ 0 & z^* {\bf 1}_n \emat.
\ee

\subsection{Integration Formula}

We have proved the following integration formula \cite{SplitVerb2}
\be
&&\int_{U(2n)} dU {\det}^\nu U e ^{\frac 12 {\rm Tr} [ M(U + U^{-1})] + 
\sum_p a_p {\rm Tr}[(UBU^{-1} B)^p ]} \nonumber\\ 
&=& \frac {c_n}{(xy)^{n(n-1)}} \det 
\left[ (x\del_x)^k (y\del_y)^l Z_{1}^\nu(x,y) \right]_{
0\le k,l \le n-1},
\label{int}
\ee
where
\be
Z^\nu_{1}(x,y) = \int_0^1 \lambda d\lambda I_\nu(\lambda x) I_\nu(-\lambda y)
e^{2\sum_p a_p \cos(2p\cos^{-1} \lambda) },
\ee
and $c_n$ is an $n$-dependent constant.
For example, consider the case $n=1$ and all $a_p =0$. Then the integral
is given by $Z_{1}(x,y)$ which is a known integral given by
\be 
\left .Z_{1}^\nu(x,y)\right |_{a_p =0} &=& \int_0^1 \lambda d\lambda 
I_\nu(\lambda x) I_\nu(-\lambda y)\nonumber \\
&=&\frac{yI_\nu(x) I_{\nu-1}(-y) + xI_{\nu-1}(x) I_{\nu}(-y)}
{x^2-y^2},
\ee
which is a known result for the QCD partition function with two different 
masses \cite{jsv}.

\subsection{Toda Lattice Equation}

Using the integration formula of previous section we find that
the low-energy limit of the phase quenched QCD partition function at
nonzero chemical potential is  given by
\be
Z_n^\nu(z,z^*) = \frac{c_n}{(zz^*)^{n(n-1)}} \det\left[(z\del_z)^k 
(z^*\del_{z^*})^l 
Z_1^\nu(z,z^*)\right]_{0\le k,l\le n-1},
\ee
where 
\be
Z_1^\nu(z,z^*) = \int_0^1 \lambda d\lambda e^{-2VF^2\mu^2 (\lambda^2-1)}
|I_\nu(\lambda zV\Sigma)|^2.
\ee 
Using the Sylvester identity as at zero chemical potential
we obtain the Toda lattice equation
\be
z\del_z z^*\del_{z^*} \log Z_n^\nu(z,z^*) =
\frac{\pi n}2 (zz^*)^2 \frac{Z_{n+1}^\nu(z,z^*) Z_{n-1}^\nu(z,z^*)}
{[Z_n^\nu(z, z^*)]^2}.
\ee
The spectral density (\ref{rhoreplica}) follows from the replica limit of
this equation. Using $Z_0^\nu(z, z^*)=1$ we find the simple expression
\be
\rho(z,z^*) = \lim_{n \to 0} \frac 1{\pi n} \del_z \del_{z^*} \log
Z_n^\nu(z, z^*) =
\frac {zz^*}2 Z_1^\nu(z,z^*) Z_{-1}^\nu(z,z^*).
\ee
What remains to be done is to calculate the partition function with one bosonic
and one conjugate bosonic quark which will be completed in the
next subsection.

\subsection{The Bosonic Partition Function}
In this subsection we evaluate the low-energy limit of the QCD
partition function at nonzero chemical potential
for one bosonic quark and one conjugate bosonic quark. Because of convergence 
requirements it is more complicated to derive the 
chiral Lagrangian in this case. By a careful analysis we 
find~{ \cite{SplitVerb2}},
\be
Z_{-1}^{\nu}(z,z^*) = \int_{U \in Gl(2)/U(2)}\hspace{-1cm} dU   {\det}^\nu U
e^{-\frac{F^2\mu^2 V}4 {\rm Tr}[U,B][U^{-1},B]
+\frac {i\Sigma V}2{\rm Tr}\zeta^T(U-I U^{-1} I)},\nonumber\\
\label{z-1} 
\ee
where $B$ is the baryon number matrix defined in (\ref{BM}) and the
mass matrix $\zeta$ and the anti-symmetric matrix $I$ are defined as
\be
\zeta = \mat \epsilon & z \\ z^* & \epsilon \emat 
\quad {\rm and} \quad 
I =\mat 0 & 1 \\ -1 & 0 \emat .
\ee
Although this integral is convergent for ${\rm Im }\epsilon > 0$, it diverges
logarithmically for $\epsilon \to 0$. We have also \cite{SplitVerb2}
derived the partition 
function (\ref{z-1}) starting from a chiral Random Matrix Theory at nonzero
chemical potential and using the Ingham-Siegel integral \cite{Fyodorov-is}.

The integral (\ref{z-1}) can be evaluated 
analytically by using an explicit parameterization of positive 
definite $2\times 2$ matrices. We find
\be
Z^\nu_{-1}(z=x+iy) = 
C_{-1}\log \epsilon\, 
e^{\frac{V\Sigma^2 (y^2-x^2)}{4 \mu^2F^2}}
K_\nu\left (  \frac {V\Sigma^2(x^2+y^2)}{4 \mu^2 F^2}\right ).
\label{z-1a}
\ee
The final result for the quenched spectral density is
given by \cite{SplitVerb2}
\be
\rho(x,y) &=& \frac {V^3\Sigma^4}{2\pi F^2\mu^2}(x^2+y^2)
e^{\frac{V\Sigma^2 (y^2-x^2)}{4 \mu^2F^2}}
K_\nu\left ( \frac {V\Sigma^2(x^2+y^2)}{4 \mu^2 F^2}\right )
\nonumber \\ && \times
\int_0^1 \lambda d\lambda e^{-2VF^2\mu^2 \lambda^2}
|I_\nu(\lambda (x+iy)V\Sigma)|^2.
\label{final}
\ee
The constant has been chosen such that    the $\mu \to 0$  limit
of $\rho(x,y)$ for large $y$ is given by $\Sigma V/\pi$ (see below).
The solid curve
in Fig. \ref{fig2} shows a graph of this result for $y=0$ 
and $\mu^2F^2 V =16$ in terms of the ratio
\be
\rho_s(x,y) = \frac {\rho(x,y)}{\Sigma^2 V^2}  
\ee
versus $x \Sigma V$ at $y=0$.
The dotted curve shows the result which 
is obtained when the Bessel function $K_\nu$ is replaced by
its asymptotic expansion. This result that was obtained from
a nonhermitian eigenvalue model \cite{akemann} that
is not in the universality class
of QCD. 
An important difference between the two results is that the spectral
density (\ref{final}) for $y=0$ is quadratic in $x$ for $ x \to 0$, whereas 
the result given by the dotted curve is linear in $x$ for $x\to 0$.

Taking the thermodynamic limit at fixed $z$ and $\mu$ the Bessel functions
can be approximated by their asymptotic limit. This results in
\be
\rho(x,y) = \frac {V\Sigma^2}{4\pi \mu^2 F^2 } \quad {\rm for} 
\quad |x |< \frac {2F^2 \mu^2}\Sigma,
\ee
and $\rho(x,y) =0 $ outside this strip in agreement with a  mean field
analysis \cite{misha,TV}.
For the integrated eigenvalue density we then find
\be
\int_{-\infty}^\infty dx \rho(x,y) = \frac{\Sigma V}\pi
\ee
in agreement with the eigenvalue density at $\mu = 0$.

\begin{figure}[ht] 
\vspace{-1.8cm}
\centerline{\hspace*{1.5cm}\epsfysize=8cm\epsfbox{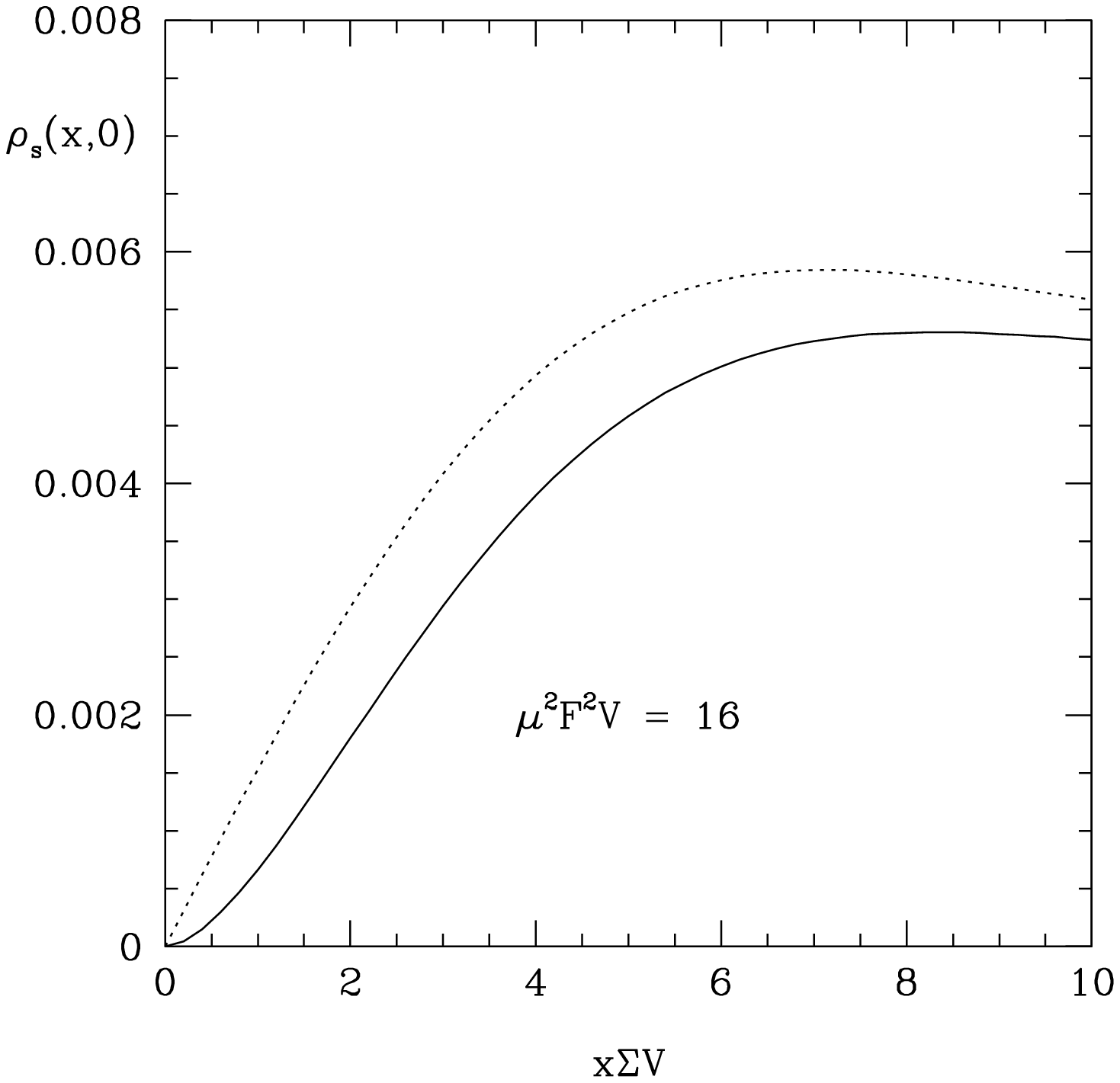}\hspace*{-2.2cm}
\epsfxsize=8cm\epsfbox{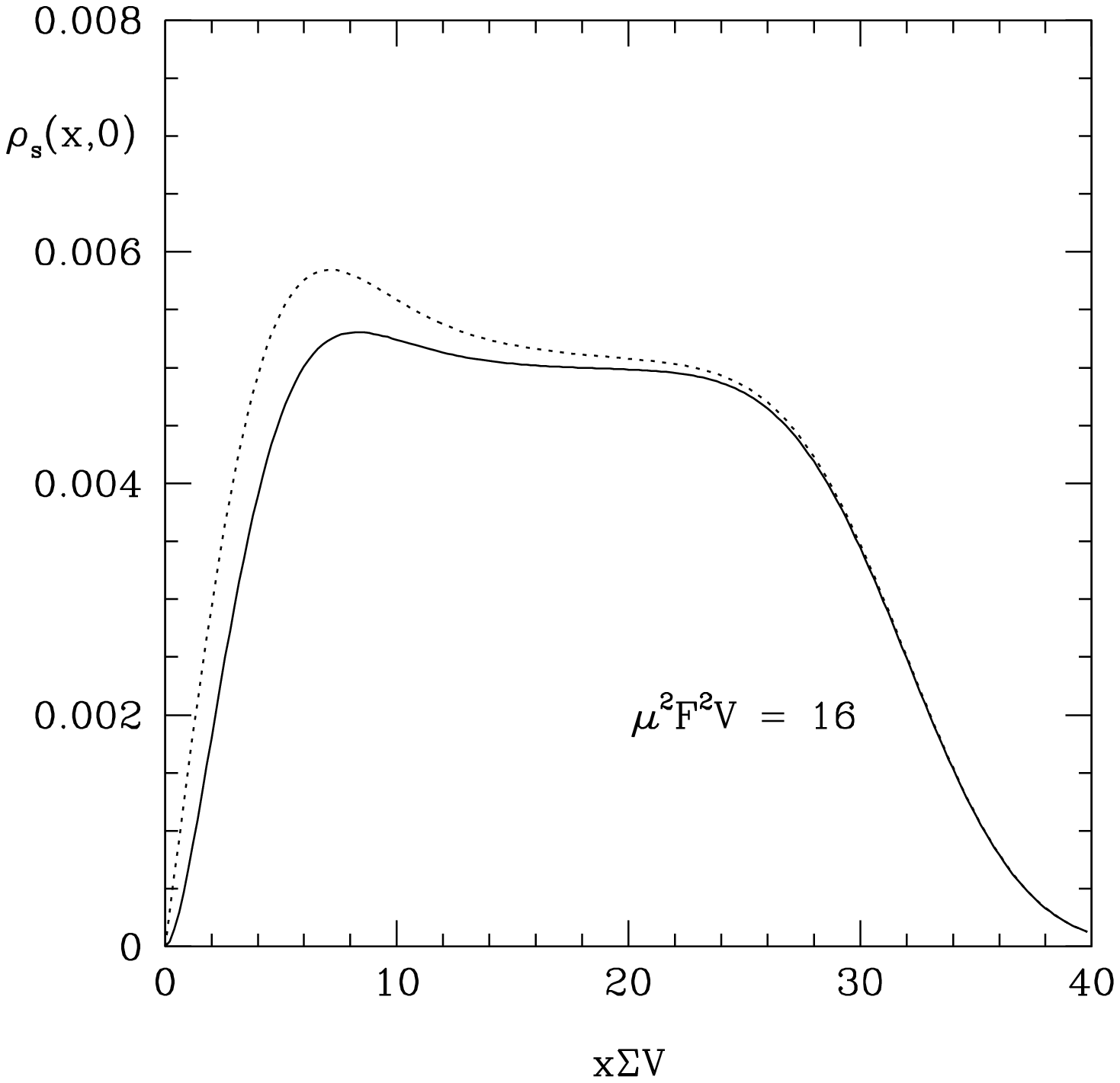} 
}   
\caption{ The quenched spectral density at nonzero chemical potential in
the ergodic domain of QCD (full curve). Also shown is a result derived
from an eigenvalue model (dotted curve). The left hand plot is a zoom in of
the right hand one. 
\label{fig2}}
\end{figure}

\section{Conclusions}
In this work we have analyzed the ergodic domain of the QCD partition 
function where the pion Compton wavelength is much larger
than the size of the box. In this domain the QCD partition function
reduces to a theory of weakly interacting Goldstone bosons for which
the kinetic term in the chiral Lagrangian can be ignored.
Independent of the quark masses, the generating function
for the Dirac spectrum is in this domain for sufficiently small
eigenvalues.

We have shown that fermionic partition functions, bosonic partition
functions and the supersymmetric partition function are connected
by a Toda lattice equation. This recursion relation  makes it possible 
to derive nonperturbative results using the replica trick. In particular, 
this reveals the factorization of the resolvent and the susceptibility
into products of simple bosonic and fermionic partition functions.
In this article, we have considered the resolvent at zero chemical
potential and the spectral density at nonzero chemical potential.

The resolvent for quenched QCD at nonzero chemical potential approaches
zero linearly as a function of ${\rm Re}(z)$. 
In the unquenched theory, where the argument of the resolvent is also 
the mass
in the fermion determinant, we expect a discontinuity in the resolvent
as ${\rm Re}(z)$ crosses the imaginary axis. Can we
understand the 
differences between these two theories in terms of the spectrum
of the Dirac operator? As is suggested by earlier Random Matrix
Theory simulations \cite{HJV}
there are significant differences between the two.
For example, in the unquenched theory, because of
the phase of the fermion determinant, there is no reason that the spectral
density is positive definite or even real.
The first analytical results for the unquenched spectral density where
recently obtained by James Osborn \cite{james}
for a nonhermitian Random Matrix
Model that is in the universality class of QCD at nonzero chemical 
potential. In the thermodynamic limit his results show strong
oscillations but the connection with broken chiral symmetry is still
a mystery. We hope to address this issue in a future 
publication \cite{AOSV}.  

The replica limit of the Toda lattice equation is a powerful 
method that is also applicable to other partition functions with
an integrable structure. For example, we mention the Ginibre ensemble
\cite{kanzieper03}, parametric correlations and the two-point function
of the Gaussian Unitary Ensemble. Our experience tells us that all
universal results that can be derived from a complex random
matrix theory ($\beta =2)$ can also be obtained from the replica
limit of a Toda lattice equation.

\section*{Acknowledgments}
We thank Gernot Akemann, Gerald Dunne and James Osborn
for useful discussions.

\end{document}